\documentstyle[12pt,aaspp4]{article}

\newcommand{\etal}{{\it et al.}}
\newcommand{\kms}{\rm~km\,s\mbox{$^{-1}$}}
\newcommand{\cc}{\rm~cm\mbox{$^{-3}$}}

\newcommand{\cit}[1]{$^{\ref{#1}}$}
\newcommand{\citt}[2]{$^{\ref{#1},\ref{#2}}$}
\newcommand{\cittt}[3]{$^{\ref{#1},\ref{#2},\ref{#3}}$}
\newcommand{\citttt}[4]{$^{\ref{#1},\ref{#2},\ref{#3},\ref{#4}}$}

\begin{document}
\vskip 0.6 in
\noindent
 
\title{\bf Coexisting conical bipolar\\
           and equatorial outflows from \\
           a high-mass protostar \\ }

\author{
L.~J. Greenhill\footnotemark[1], 
C.~R. Gwinn\footnotemark[2],
C. Schwartz\footnotemark[3],
J.~M. Moran\footnotemark[1],
P.~J. Diamond\footnotemark[4]
}

\footnotetext[1]{Center for Astrophysics, 60 Garden Street, Cambridge, MA 02138.}
\footnotetext[2]{University of California, Santa Barbara, CA 93106.}
\footnotetext[3]{Maria Mitchell Observatory, Nantucket, MA 02554. 
                 Currently Colby College, Waterville, ME 04901.}
\footnotetext[4]{National Radio Astronomy Observatory, PO Box O, Socorro, NM 87801.}

\slugcomment{**** Accepted by Nature 11/19/98 ****}
 
{\bf 
The BN/KL region in the Orion molecular cloud\cit{Kleinmann67} is an archetype
in the study of the formation of stars much more massive than the
Sun\cit{Genzel89}. This region contains luminous young stars and protostars,
but it is difficult to study because of overlying dust and gas. Our basic
expectations are shaped to some extent by the present theoretical picture of
star formation, the cornerstone of which is that protostars acrete gas from
rotating equatorial disks, and shed angular momentum by ejecting gas in
bipolar outflows. The main source of the outflow in the BN/KL region
\cittt{Gezari92}{Gezari95}{Menten95} may be an object known as radio source
I\cit{Churchwell87}, which is commonly believed to be surrounded by a rotating
disk of molecular material \cittt{Plambeck90}{Wright95}{Barvainis84}. Here we
report high-resolution observations of silicon monoxide (SiO) and water maser
emission from the gas surrounding source I; we show that within 60 AU (about
the size of the Solar System), the region is dominated by a conical bipolar
outflow, rather than the expected disk. A slower outflow, close to the
equatorial plane of the protostellar system, extends to radii of 1,000 AU.
}

The BN/KL region radiates an estimated 5--8$\times10^4$ erg\,s$^{-1}$
\citt{Gezari95}{Thronson86} and is a rich source of molecular line
emission\citt{Genzel89}{Wright96} embedded within the Orion Molecular Cloud
(OMC-1). Two molecular outflows originate in BN/KL, relatively close to source
{\bf I}, a low-velocity (18\kms) outflow to the
northeast-southwest\citt{Wright96}{Genzel81}, and an orthogonal high-velocity
(30--100\kms) outflow\citttt{Wright96}{Genzel81}{Allen93}{Jones85}, both of
which extend over about $1'$ (0.14 pc). Source {\bf I} has a broad-band radio
spectrum consistent with an H\,II region about 60 AU in
diameter\cit{Plambeck95}, though it has not yet been specifically detected at
infrared wavelengths. Masers are often associated with massive young stars,
and many are broadly distributed across BN/KL (masers are the microwave
equivalent of lasers). However, a distinct population of SiO and H$_2$O masers
is clustered around source {\bf I}\citt{Genzel89}{Gezari92}. To
understand  better the physical nature of source {\bf I} and its relationship to the
outflows in BN/KL, we have mapped these masers with high spectral and angular
resolution. 
We resolve the angular structure of the SiO emitting region in detail at
each observed line-of-sight velocity, in contrast to previous observations
with connected-element radio 
interferometers\citttt{Menten95}{Plambeck90}{Wright95}{Morita92}.

Maser sources are valuable tracers in molecule-rich regions because they
comprise many compact, high surface-brightness, spatially distinct points of 
radiation (spots), with various line-of-sight velocities. The spots are 
test ``particles'' tracing the three-dimensional velocity field of
the material in which they are embedded. Maser amplification requires
relatively small gradients in line-of-sight velocity to achieve long gain-path lengths.  To
pump the maser levels by collisions, the gas temperature must also be 
at least a
few hundred degrees, and the density orders of magnitude greater 
than that sampled by
thermal molecular emission, i.e., n$_{H_2}\ga 10^8\cc$.  Maser emission is a 
selective tracer but in star forming regions, it can in principle occur in either
outflows or disks\cit{Torrelles98}. 

We observed emission from the $v=1$, $J=$1--0 transition of SiO (43.12208 GHz
rest frequency) toward the BN/KL region on 9 July 1995 with the Very Long
Baseline Array (VLBA) of the NRAO.  The data were processed with the VLBA
correlator, which provided 0.22\kms~wide spectral channels, and reduced with
standard techniques\cit{Zensus95}. Within 30\kms~of the 5\kms~systemic
velocity\cit{Genzel89} of the region near source {\bf I}, we detected 1,212 maser spots
with flux densities between 0.09 and 300 Jy.

Spectroscopic images show that the SiO maser emission is confined to four
regions that can be envisioned as marking the arms of an {\bf X}, subtending
about $0\rlap{.}''25$ (Figure\,1a). We propose that the {\bf X} marks the
intersection of two outflow cones with the plane of the sky and separates the
low-velocity (northeast-southwest) and high-velocity (northwest-southeast)
outflows in BN/KL. Source {\bf I} was not detected by us because its surface
brightness is too low.  However, to estimate the position of source {\bf I} on
our maps, we compared the flux-weighted emission centroid for each VLBA
spectral channel to the earlier, lower resolution ($0\rlap{.}''25$) map  
of Menten \& Reid\cit{Menten95} who
measured the position of source {\bf I} relative to the maser emission
centroids with $0\rlap{.}''01$ uncertainty, using the Very Large Array (VLA).
The relative positions of the dominant clumps of masers in the
two maps agree to $0\rlap{.}''01$. Source {\bf I} lies at the center
of the {\bf X}, within the measurement errors (Figure\,1a). The innermost SiO
maser features are separated by about the 60~AU diameter of the 
source {\bf I} H\,II region.

We have added to our study archived H$_2$O maser data collected 
on 25 August 1983 with the VLA (Figure\,1b),
as well as published maps\cit{Gaume98} of the H$_2$O maser distribution on 5
August 1991. The distribution of the H$_2$O masers is elongated in the
direction of the low-velocity outflow in BN/KL. Over eight years,
the length of the distribution grew by about $0\rlap{.}''2$.  We measured
proper motions for 19 individual maser features that persisted over the eight
year period (Figure\,1b). The motions are primarily parallel to the axis
of the low-velocity outflow, and the average transverse motion corresponds to 
about 20\kms, approximately the speed of the large scale low-velocity
outflow in BN/KL\cit{Genzel81}.

The previous widely accepted model, in which the SiO masers lie in a rotating,
expanding, and inclined disk\cittt{Plambeck90}{Wright95}{Barvainis84}, does not
fit our high-resolution VLBA maps. One attraction of this model
was that the disk could feasibly extend to larger radii, explaining the
low-velocity BN/KL outflow. In such a disk, the SiO emission is expected to trace dynamic surfaces where the
derivative of the line-of-sight velocity is small, so that the maser
gain is high.  For most rotation laws, each arm of the {\bf X} would
be expected to show a velocity gradient along its length; but three of four do not. 
Moreover, the distribution of SiO masers that we observe shows approximate
mirror symmetry in position and velocity about a northwest-southeast axis,
which constrains the magnitude of any rotation to be less than a few \kms. If
the disk is gravitationally bound to the protostar, then this rotation speed,
at 25--60 AU radius, constrains the protostellar mass to be less than 1
M$_\odot$. If radial forcing enlarges the disk (e.g., a wind) from an initially
bound state (with about 1 AU radius), then conservation of angular momentum
requres a stellar mass less than a few M$_\odot$. These relatively small
stellar masses are inconsistent with the presence of an H\,II
region.

We propose that the SiO maser emission outlines the limbs of a bipolar,
conical, high-velocity flow 25--60 AU from the protostar (i.e., the {\bf X}), and
that the H$_2$O masers lie in a low-velocity flow in an equatorial band
bisecting the high-velocity flow (Figure\,1c). The position angles of the cones
are $-40^\circ$ and $150^\circ$, with opening angles of $60^\circ$ and
$80^\circ$, respectively. For an outflow axis close to the plane of the sky
and a finite cone wall thickness, the longest maser gain paths naturally lie
along the limbs of the cones.  The  10--15\kms~difference in
mean Doppler velocity between the red- and blue-shifted SiO masers may be
explained if the cones are tipped out of the plane of the sky. The
known distribution of H$_2$ emission in BN/KL\cit{Sugai95} shows two lobes
with line-of-sight velocities Doppler shifted by more than 
30\kms~that subtend the openings of the
cones. The northwest lobe is equally prominent in red- and blue-shifted
emission, which suggests that the inclination is small. (Blue-shifted emission from the
southeast lobe appears to be blocked by a dense foreground
cloud\citt{Plambeck95}{Migenes89}.)

The dense gas in which the SiO masers lie is probably not gravitationally bound
to the protostar because there is no detectable rotation. The SiO maser
emission probably arises at a steady-state shock interface, perhaps between a
faster and a slower component of the stellar outflow. Such shocks serve as
efficient mechanisms that collisionally pump SiO masers and reduce depletion
of gas-phase SiO onto dust grains. To quantify the lifetime of the visible
structure, we compared the present SiO maser maps with a partially complete
VLBI map obtained in 1985\cit{Greenhill88}.  The loci of red-shifted masers
have remained fixed with respect to source {\bf I} to about
$0\rlap{.}''01$ over 10.3 years, which corresponds to a proper motion of less
than 4.5 AU or 2\kms. We speculate that collimation of the bipolar outflow
begins at radii less than 25 AU, perhaps inside the H\,II region, because
source {\bf I} lies close to the apices of the outflow cones.

We fit the H$_2$O maser positions and Doppler velocities to a
constant-velocity outflow centered on source {\bf I}, where the H$_2$O masers
are constrained to lie on a toroidal surface, as suggested
qualitatively by the curvature traced by several H$_2$O features (Figure\,1c).
This surface may be where large-diameter dust grains begin to form and to be
accelerated by radiation pressure. (Dust condensation and H$_2$O maser
excitation together require temperatures of 400--1,000 K.  If this material is
in thermal equilibrium, then the luminosity of source {\bf I} would be $\gg
10^4$ L$_\odot$.) The torus has the following parameters: mean outer radius 
360 AU, thickness 270--320 AU, outflow velocity 13\kms, and rotation less 
than a few \kms. The torus is tipped down about
$10^\circ$ with respect to the line of sight, as evidenced by the southeasterly position bias of the distribution of
blue-shifted H$_2$O masers. (The masers lie close to the equator.)  
We assume that the axes of the high-velocity
outflow and the torus are aligned (Figure\,1c), from which we infer
that the high-velocity outflow is at least 50\kms.  This limit is consistent
with the observed three-dimensional motions of Herbig-Haro objects in
BN/KL\citt{Hu96}{Jones85} and the Doppler velocities of thermal CO and SiO
emission\cit{Wright96}. We note that the SiO maser emission from the limbs of
the cones suggests material flowing away from the observer toward the
northwest, opposite that expected given inclination of the torus.  However, we
consider that the masers trace gas selectively. For small inclinations,
optimal amplification can occur with the opposite sense 
of Doppler shift (i.e., red- vs blue-shift) from gas along the axes of the cones,
depending on the radial and angular dependence of the outflow
velocity.

We conclude that the equatorial region is dominated by
outflowing material and not by (rotating) natal material.   We may extend this
outflow to radii of up to 1,000 AU by considering the previously mapped
distribution of v$=$0 SiO emission\citt{Wright95}{Chandler95}, to which three
components contribute: high-velocity thermal, low-velocity thermal,  and
low-velocity maser emission. Both low-velocity components overlie the H$_2$O
maser distribution in velocity and sky position within 350 AU of source {\bf I}.
The brightness distribution is asymmetric with respect to source {\bf I}, as is
the H$_2$O maser torus.  On the smaller side of the
torus, stronger SiO emission coincides with a richer distribution of H$_2$O 
masers, which is consistent with a density gradient in the surrounding medium,
as seen at 8.7 and 12.4 $\mu$m\citt{Gezari92}{Gezari95}. Although the apparent
mixing of H$_2$O maser and SiO emission is difficult to understand (since
silicate-grain formation depletes gas-phase SiO), we note that similar mixing
is apparent in the envelopes of late-type stars\citt{Lucas92}{Danchi94}.

We note three propitious alignments of structure that strengthen the case for
the proposed model.   First, no H$_2$O masers lie within the high-velocity
flow demarcated by the SiO masers (Figure\,1c). This flow is probably
too hot or turbulent to support maser action. Second, the four components of
IRc2\cit{Dougados93} shown in Figure\,2 subtend the opening angle of the
northern outflow cone. They may have a direct line of sight to the protostar
and may be heated by it or by interaction with the high-velocity flow.
Infrared light from the southeastern outflow is blocked by extinction in a
foreground cloud core\citt{Plambeck95}{Migenes89}. Third, the magnetic field
direction inferred from dust emission at 100 $\mu$m and 3 mm wavelength is
parallel to the high-velocity outflow axis\citt{Schleuning98}{Rao98}. This
suggests that the formation of the protostar and the alignment of the outflow
axis may have been affected by the broad field distribution in OMC-1.

We argue that the masers clustered around source {\bf I} provide evidence for
a two-component outflow from a young star. In the context of star formation
theory for high-mass stars, no current model readily explains wide-angle
collimation of a high-velocity bipolar outflow within tens of AU of a
protostar and a simultaneous low-velocity equatorial outflow.
Magnetohydrodynamic (MHD) mechanisms such as X-winds\cit{Shu95} and disk
winds\cit{Wardle93} may be reasonable candidates. Alternatively, a spherical
stellar wind that mixes with and ablates an (unseen) accretion disk up to about 
350 AU in radius might
generate a slow, dense equatorial wind, but this would not explain the
observed conical structures.  In addition, the presence of an H\,II region may
preclude a disk. If accretion and outflow processes are linked, then
the birth of an H\,II region could dissipate the observed equatorial outflow in 
about 100--200 years. Based on the alignment of the outflows from source {\bf
I} with other structures in BN/KL, we suggest that source {\bf I} drives a
substantial portion of the bulk motion in BN/KL, though other sources,
such as {\bf n} (Figure\,2), rightly draw attention\citt{Gezari95}{Menten95}. 
We anticipate that VLBA maps of SiO and
H$_2$O maser proper motions and SiO maser polarizations will clarify the
structure, dynamics, and magnetic field geometry in this region, which will
help resolve the debate about sources of energy and
outflow in BN/KL.

\centerline{\bf REFERENCES}

\begin{enumerate}
\item \label{Kleinmann67}
  Kleinmann, D.~E., \& Low, F.~J.
  Discovery of an infrared nebula in Orion.
  {\sl Astrophys. J.} {\bf 149}, L1--L4 (1967).

\item \label{Genzel89}
  Genzel, R. \& Stutzki, J.
  The Orion molecular cloud and star-forming region.
  {\sl Ann. Rev. Astron. Astrophys.} {\bf 27}, 41--85 (1989).

\item \label{Gezari92}
  Gezari, D.~Y.
  Mid-infrared imaging of Orion BN/KL: astrometry of IRc2 and the SiO maser.
  {\sl Astrophys. J.} {\bf 396}, L43--L47 (1992).

\item \label{Gezari95}
  Gezari, D.~Y. \& Backman, D.~E.
  4.8--20 micron imaging of Orion BN/KL: II. A new look at luminosity
    sources and the role of IRc2.
  {\sl Astroph. \& Space Sci.} {\bf 224}, 45--52 (1995).

\item \label{Menten95}
  Menten, K.~M. \& Reid, M.~J.
  What is powering the Orion Kleinmann-Low infrared nebula?
  {\sl Astrophys. J.} {\bf 445}, L157--L160 (1995).

\item \label{Churchwell87}
  Churchwell, E., Wood, D.~O.~S., Felli, M., \& Massi, M.
  Solar system-sized condensations in the Orion Nebula.
  {\sl Astrophys J.} {\bf 321}, 516--519 (1987).

\item \label{Plambeck90}
  Plambeck, R.~L., Wright, M.~C.~H. \& Carlstrom, J.~E.
  Velocity structure of the Orion-IRc2 SiO maser: evidence for an 80~AU diameter
    circumstellar disk.
  {\sl Astrophys. J.} {\bf 348}, L65--L68 (1990).

\item \label{Wright95}
  Wright, M.~C.~H., Plambeck, R.~L., Mundy, L.~G. \& Looney, L.~W.
  SiO emission from a 1,000~AU disk in Orion KL.
  {\sl Astrophys. J.} {\bf 455}, L185--L188 (1995).

\item \label{Barvainis84}
  Barvainis, R.
  The polarization of the SiO masers in Orion: maser emission from a rotating,
    expanding disk?
  {\sl Astrophys. J.} {\bf 279}, 358--362 (1984).

\item \label{Thronson86}
  Thronson, H.~A.~Jr., \etal~
  The Orion star-forming region: far-infrared and radio molecular 
  observations.
  {\sl Astron. J.} {\bf 91}, 1350--1356 (1986).

\item \label{Wright96}
  Wright, M.~C.~H., Plambeck, R.~L. \& Wilner, D.~J.
  A multiline aperture synthesis study of Orion-KL.
  {\sl Astrophys. J.} {\bf 469}, 216--237 (1996).

\item \label{Genzel81}
  Genzel, R., Reid, M.~J., Moran, J.~M. \& Downes, D.
  Proper motions and distances of H$_2$O maser sources. I. The outflow in Orion-KL.
  {\sl Astrophys. J.} {\bf 244}, 884--902 (1981).

\item \label{Allen93}
  Allen, D.~A. \& Burton, M.~G.
  Explosive ejection of matter associated with star formation in the Orion nebula.
  {\sl Nature} {\bf 363}, 54--56 (1993).

\item \label{Jones85}
  Jones, B.~F. \& Walker, M.~F.
  Proper motions of Herbig-Haro objects. VI. The M42 HH objects.
  {\sl Astron. J.} {\bf 90}, 1320--1323 (1985).

\item \label{Plambeck95}
  Plambeck, R.~L., Wright, M.~C.~H., Mundy, L.~G., \& Looney, L.~W.
  Subarcsecond-resolution 86 GHz continuum maps of Orion KL.
  {\sl Astrophys J.} {\bf 455}, L189--L192 (1995).


\item \label{Morita92}
  Morita, K.-I., Hasegawa, T., Ukita, N., Okumura, S.~K. \& Ishiguro, M.
  Accurate positions of SiO masers in active star forming regions:
    Orion-KL, W51-IRS2, and Sgr-B2 MD5.
  {\sl Pub. Astron. Soc. Japan} {\bf 44}, 373--380 (1992).

\item \label{Torrelles98}
  Torrelles, J.~M., et al. 
  Radio Continuum-H2O Maser Systems in NGC 2071: H2O Masers Tracing a Jet 
  (IRS 1) and a Rotating Proto-planetary Disk of Radius 20 AU (IRS 3).
  {\sl Astrophys. J.} {\bf 505}, 756-765 (1998).

\item \label{Zensus95}
  Zensus, J.~A., Diamond, P.~J. \& Napier, P.~J., eds.
  {\sl Very Long Baseline Interferometry and the VLBA}
  (Astron. Soc. Pacific, San Francisco, 1995).

\item \label{Gaume98}
  Gaume, R.~A., Wilson, T.~L., Vrba, F.~J., Johnston, K.~J. \& Schmid-Burgk, J.
  Water masers in Orion.
 {\sl Astrophys. J.} {\bf 493}, 940--949 (1998).

\item \label{Sugai95}
  Sugai, H., Kawabata, H., Usuda, T., Inoue, M.~Y., Kataza, H. \& Tanaka, M.
  Velocity field of the Orion-KL region in molecular Hydrogen emission.
  {\sl Astrophys. J.} {\bf 442}, 674--678 (1995).

\item \label{Migenes89}
  Migenes, V., Johnston, K.~J., Pauls, T.~A. \& Wilson, T.~L.
  The distribution and kinematics of ammonia in the Orion-KL nebula:
    high-sensitivity VLA maps of the NH$_3$(3,2) line.
  {\sl Astrophys. J.} {\bf 347}, 294--301 (1989).

\item \label{Greenhill88}
  Greenhill, L.~J. {\sl et al}.
  in {\sl The Impact of VLBI on Astrophysics and Geophysics}
  (eds Reid, M.~J. \& Moran, J.~M.) 253--254
  (Kluwer, Dordrecht, 1988).

\item \label{Hu96}
  Hu, X.
  Kinematic Studies of Herbig-Haro Objects in the Orion Nebula.
  {\sl Astron. J.} {\bf 112} 2712-2717 (1996).

\item \label{Chandler95}
  Chandler, C.~J. \& De Pree, C.~G.
  Vibrational ground-state SiO {$J=$1--0} emission in Orion IRc2 imaged
    with the VLA.
  {\sl Astrophys. J.} {\bf 455}, L67--L71 (1995).

\item \label{Lucas92}
  Lucas, R., \etal~
  Interferometric observations of SiO $v=0$ thermal emission from evolved stars.  
  {\sl Astron. \& Astrophys.} {\bf 262}, 491--500 (1992).

\item \label{Danchi94}
  Danchi, W.~C., Bester, M., Degiacomi, C.~G., Greenhill, L.~J., \& Townes, C.~H.
  Characteristics of dust shells around 13 late-type stars.
  {\sl Astron. J.} {\bf 107}, 1469--1513 (1994).

\item
  \label{Dougados93}
  Dougados, C., Lena, P., Ridgway, S.~T., Christou, J.~C. \& Probst, R.~G.
  Near-infrared imaging of the Becklin-Neugebauer-IRc2 region in Orion
    with subarcsecond resolution.
  {\sl Astrophys. J.} {\bf 406}, 112--121 (1993).

\item \label{Schleuning98}
  Schleuning, D.~A.
  Far-infrared and submillimeter polarization of OMC-1: evidence for magnetically
    regulated star formation.
  {\sl Astrophys. J.} {\bf 493}, 811--825 (1998).

\item \label{Rao98}
  Rao, R., Crutcher, R.~M., Plambeck, R.~L. \& Wright, M.~C.~H.
  High-resolution millimeter-wave mapping of linearly polarized dust emission:
    magnetic field structure in Orion.
  {\sl Astrophys. J.} {\bf 502}, L75--L78 (1998).

\item \label{Shu95}
  Shu, F.~H., Najita, J., Ostriker, E.~C. \& Shang, H.
  Magnetocentrifugally driven flows from young stars and disks. V. Asymptotic
    collimation into jets.
  {\sl Astrophys. J.} {\bf 455}, L155--L158 (1995).

\item \label{Wardle93}
  Wardle, M., \& K\"onigl, A. 
  The structure of protostellar accretion disks and the origin of bipolar flows.
  {\sl Astrophys. J.} {\bf 410}, 218--238 (1993).

\item \label{Garay89}
  Garay, G., Moran, J.~M. \& Haschick, A.~D.
  The Orion-KL super water maser.
  {\sl Astrophys. J.} {\bf 338}, 244--261 (1989).

\end{enumerate}

\acknowledgements
{\small We thank Mark Reid for assistance in recovering his 1983 VLA data. We
thank Catherine Dougados and Karl Menten for permission to use the infrared
and radio images presented in Figure\,2. CS was supported by the NSF REU
program. C.R.G. acknowledges support of the National Science Foundation.
The National Radio Astronomy Observatory is a facility of the U.S.
National Science Foundation operated under cooperative agreement by Associated
Universities, Inc.
}

\newpage

\centerline{ {\bf CAPTIONS} }

{\bf Figure\,1a}~The $v=1$, $J=$1--0 SiO maser distribution in the BN/KL region,
obtained with the VLBA. Unlike previous maps made with connected-element radio 
interferometers, this one 
resolves the angular
structure of the SiO emitting region at each observed line-of-sight velocity.
The cross represents the centroid and position uncertainty for continuum source
{\bf I} with respect to the masers. The double arrows indicate the direction of
the proposed high-velocity outflow from source {\bf I}, and the approximate
direction of the high-velocity outflow in BN/KL. Symbol color indicates
Doppler velocity in \kms~with respect to the Local Standard of Rest, as
indicated by the colored bar. The dark bar indicates 25~AU, for an assumed
distance of 450~pc. The interferometer beam is $0.40\times0.16$
milliarcseconds ($0.18\times0.07$~AU) and the relative position uncertainties
for the masers are smaller than the plotted symbols. We referenced the data to
the strong emission at -3.4\kms~(at the origin) to stabilize the VLBA
against atmospheric fluctuations along the line of sight. Deconvolution of the
instrumental response required special attention because most images contained
complex structure, and because of the proximity of Orion to the celestial
equator. From our VLBA data, we estimate that the origin of the map is 
$\alpha_{2000}=5^h35^m14\rlap{.}^s5101\pm0.0003$,
$\delta_{2000}=-05^\circ22'30\rlap{.}''63\pm0.05$.

{\bf 1b}~The distribution of H$_2$O masers in BN/KL within $2''$ of source {\bf I},
observed with the VLA in 1983, at $0\rlap{.}''1$ resolution. The map of the
SiO maser emission is shown within the dashed contour. The cross marks the
position of source {\bf I}. The registration uncertainty between the H$_2$O
and SiO distributions is about $0\rlap{.}''05$, limited by uncertainty in
the astrometric
position of each with respect to the extragalactic radio reference
frame.  Arrows attached to the H$_2$O
maser spots show the estimated relative proper motions over 8 years.  The mean
motion has been subtracted to account for motion of the reference emission.
The double arrows indicate the direction of the proposed high-velocity
outflow from source {\bf I}. The scale bar indicates 100 AU. An arrow as long as the bar
corresponds to a motion of about 50\kms. The 1983 VLA observations covered a
total velocity range of $-12.4$ to 28.1\kms, except for velocities close to a
$4\times10^6$ Jy H$_2$O maser flare\cit{Garay89} with a velocity of 8\kms~and
a position of $\alpha_{2000}=5^h35^m14\rlap{.}^s121\pm0.003$,
$\delta_{2000}=-05^\circ22'36\rlap{.}''27\pm0.05$. The maser data were
self-calibrated and relative positions for the maser spots measured. Absolute
positions were obtained from low dynamic range images obtained
contemporaneously by a wide-band antenna subarray that observed the maser
flare and a phase calibrator, 0529+075.

{\bf 1c}~A proposed model for the region surrounding source {\bf I}, showing
the high-velocity and low-velocity outflows schematically as cones and a
non-axisymmetric torus, respectively. The front surface of the torus is
cut-away for clarity, and the SiO and H$_2$O maser distributions are
superposed.   The arrows indicate schematically the high-velocity outflow; the
arrows at position angles of $-60^\circ$ and $120^\circ$ also correspond
approximately to the direction of the magnetic field in
OMC-1\cit{Schleuning98}.  A toroidal geometry is suggestive but not unique.
The H$_2$O masers lie close to the surface of the torus, probably in
outward-propagating shocks. Overall, the zone in which the H$_2$O masers lie
may not move significantly over time, as individual shocks move outward and
diminish over time, while new ones are born closer to the protostar.
Otherwise, the dynamical age of the equatorial outflow 
is on the order of 100 years.

{\bf Figure\,2}~ A near-infrared image (3.8 $\mu$m) of the region near
IRc2\cit{Dougados93}, with contours representing 8.4 GHz
emission\cit{Menten95} and the H$_2$O maser distribution superposed. The bar
indicates 1,000 AU. Labels mark the four components of IRc2 (A, B, C, and D)
and other infrared sources.  Specifically, source {\bf n} is the brightest
infrared object in the field. It is probably a young massive star with an
associated H\,II region\cit{Menten95}, though no dust emission have been
detected yet at 3 mm wavelength\cit{Plambeck95}. The dashed box indicates the
field of view in Figure\,1b. The {\bf X} highlights a possible relationship
bewteen the proposed outflow from source {\bf I} and IRc2. Also shown is the
center of expansion (C.O.E.) for the low-velocity (18\kms) flow in BN/KL, 
estimated in part from the proper motions of H$_2$O masers outside the field 
shown\cit{Genzel81}.

\end{document}